\documentclass{PoS}

\usepackage{color}

\newcommand{\be}{\begin{eqnarray}}
\newcommand{\ee}{\end{eqnarray}}

\newcommand{\pd}{\partial}

\title{Application of the Lefschetz thimble formulation to the (0+1) dim.\ Thirring model at finite density}

\ShortTitle{Lefschetz thimble formulation to the (0+1) dim.\ Thirring model at finite density}

\author{Hirotsugu Fujii\\
        Institute of Physics, University of Tokyo, Tokyo 153-8092, Japan\\
        E-mail: \email{kikukawa@hep1.c.u-tokyo.ac.jp}}

\author{\speaker{Syo Kamata}\\%
        Rikkyo University, Tokyo 171-8501, Japan\\
        E-mail: \email{skamata@rikkyo.ac.jp}}

\author{Yoshio Kikukawa\\
        Institute of Physics, University of Tokyo, Tokyo 153-8092, Japan\\
        E-mail: \email{kikukawa@hep1.c.u-tokyo.ac.jp}}

\abstract{
Based on the Lefschetz thimble formulation of path-integration,
we analyze the (0+1) dimensional Thirring model at finite chemical potentials
and perform hybrid Monte Carlo (HMC) simulations. 
We adopt the lattice action defined with the staggered fermion and
a compact link field for the auxiliary vector field. 
We firstly locate the critical points (saddle points) of the gradient flows
within the subspace of time-independent (complex) link field,
and study the thiemble structure and the Stokes phenomenon 
to identify the thimbles which contribute to the path-integral. 
Then, we perform HMC simulations on the single dominant thimble and compare the results to the exact solution.
The numerical results are in agreement with the exact ones in small and large chemical potential regions, 
while they show some deviation in the crossover region in the chemical potential.
We also comment on the necessity of the contributions from multiple thimbles in the crossover region.
}

\FullConference{The 33rd International Symposium on Lattice Field Theory\\
                 14 -18 July  2015\\
                 Kobe International Conference Center, Kobe, Japan}

\begin{document}

\section{Introduction}
Exploring properties of QCD at finite temperatures and densities is one of the exciting topics in particle physics.
Lattice QCD has proved to be a powerful nonperturbative method at finite temperatures,
but its direct application at finite densities is extremely hard because of the sign problem;
for nonzero chemical potentials, the action becomes complex and
the integrand becomes a highly oscillatory function, which invalidates the importance sampling.
Usual Monte Carlo algorithms break down and we can not extract information of high density region of QCD.
Various methods have been applied for studying QCD at finite density, which include, e.g.,
Taylor expansion method, multi-reweighting method, use of imaginary chemical potential, study of SU(2) theory, complex Langevin method 
and so on\cite{deForcrand:2010ys, Sexty:2014dxa}. 

In this work, we try to apply the Lefschetz thimble method\cite{Witten:2010cx,Cristoforetti:2012su,Fujii:2013sra} to the lattice (0+1) dimensional Thirring model at finite density,  which is proposed in Ref.~\cite{Pawlowski:2014ada}.
This model has several similarities to lattice QCD at finite chemical potentials:
(1) it is described by compact variables, 
(2) the sign problem occurs from its Dirac determinant, 
(3) the fermion number density and the chiral condensate are physically important observables.
Moreover, differently from QCD, 
(4) the partition function of this model can be evaluated exactly.
We firstly locate the critical points (saddle points) of the gradient flows within the subspace of time-independent (complex) link field, and study the thiemble structure and the Stokes phenomenon to identify the thimbles which contribute to the path-integral. We next perform HMC simulations on the single dominant thimble and compare the results to the exact solution.
This sutdy of the Thirring model with the thimble method will give us 
useful lessons for its future application to the (3+1) dimensional QCD at finite density.

\section{Model definition and application of the Lefschetz thimble method}
In this investigation, we use the lattice action\cite{Pawlowski:2014ada}
\be
&& S=\sum_{n=1}^{L} \beta N_{f} (1-\cos A_{n}) - \sum_{i=1}^{N_{f}} \log \det D_{i}, \\
&& D_{i\, nm} = \frac{1}{2} \left[ e^{\mu_{i} +iA_{n}} \delta_{n+1,m}  - e^{-\mu_{i} +iA_{m}} \delta_{n,m+1} \right] + m_{i} \delta_{n,m},
\ee
where $L$ is the lattice size, $N_{f}$ is the flavor number, $\beta$ is the inverse coupling squared $\beta=1/2g^{2}$, $m_{i}$ is the fermion mass, and $\mu_{i}$ is the chemical potential.
The auxiliary field $A_{n}$ is defined as a compact link-auxiliary field $e^{i A_{n}}$ on links,
and takes the values in the range, $-\pi \le A_{n}< \pi$.
In our study, we focus only on the single flavor, $N_{f}=1$.
The partition function can be evaluated exactly,
\be
&& Z= \frac{e^{-L\beta}}{2^{L-1}} \left[  I_{1}(\beta)^{L} \cosh L \mu  + I_{0}(\beta)^{L} \cosh L\hat{m} \right], \qquad \hat{m} =\sinh^{-1} m ,
\ee
where $I_{0}(x)$ and $I_{1}(x)$ are the modified Bessel functions.
From this exact form, we can obtain the exact number density and chiral condensate by differentiating the partition function with respect to the chemical potential and the fermion mass, respectively,
\be
&& \langle n \rangle \equiv \frac{1}{L} \frac{\pd \log Z}{\pd \mu} = \frac{I_{1}(\beta)^{L} \sinh L \mu }{I_{1}(\beta)^{L} \cosh L \mu + I_{0}(\beta)^{L} \cosh L \hat{m}}, \\
&& \langle \bar{\chi} \chi \rangle \equiv \frac{1}{L} \frac{\pd \log Z}{\pd m} = \frac{I_{1}(\beta)^{L} \sinh L \mu }{ \left[ I_{1}(\beta)^{L} \cosh L \mu + I_{0}(\beta)^{L} \cosh L \hat{m}\right] \cosh \hat{m}}.
\ee

\begin{figure}[thbp]
  \begin{center}
    \begin{tabular}{cc}
      \begin{minipage}{0.5\hsize}
        \begin{center}
          \includegraphics[clip, width=50mm]{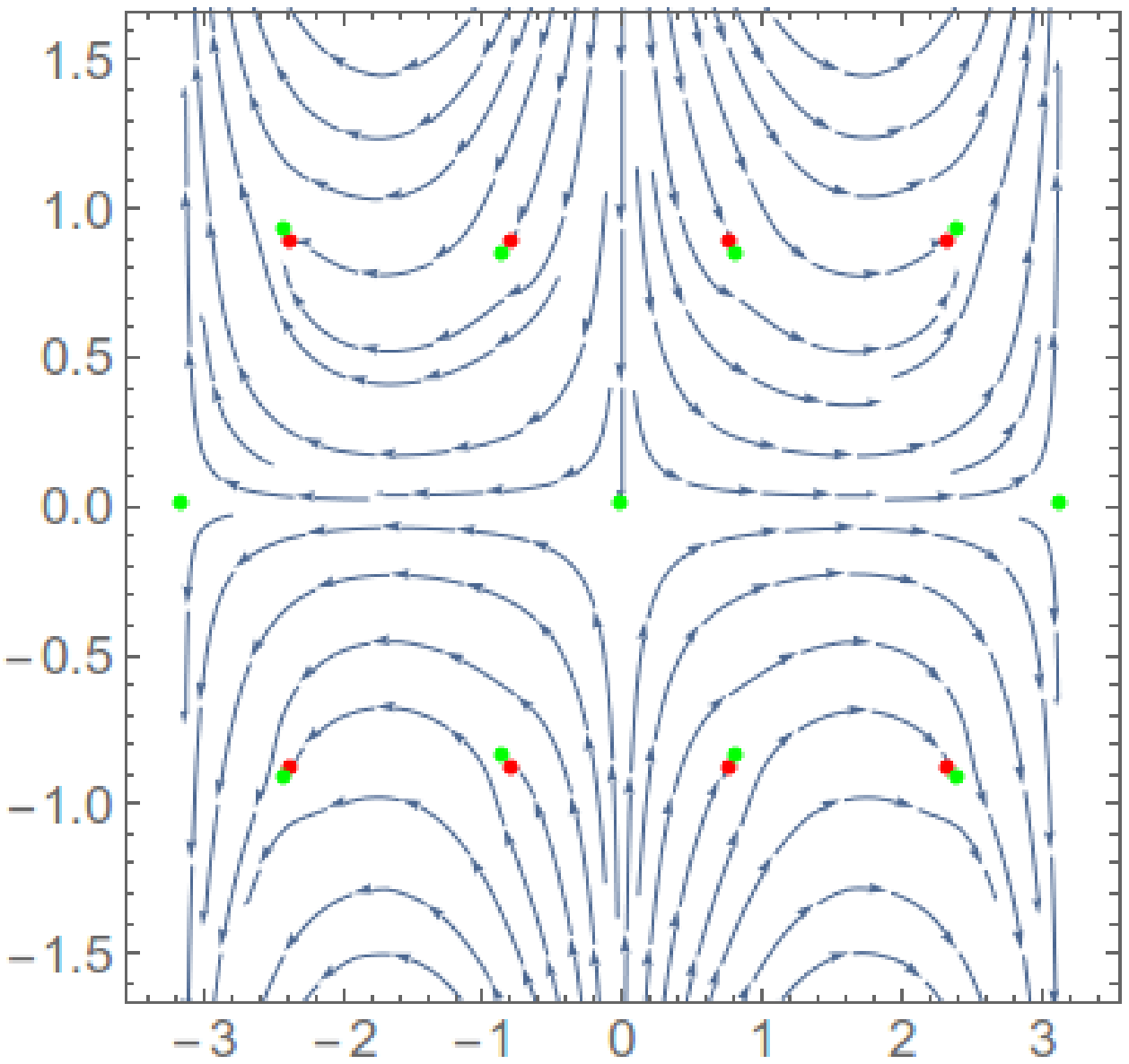}
          \hspace{1.6cm} \\(a) $\mu=0$
        \end{center}
      \end{minipage} 
      \begin{minipage}{0.5\hsize}
        \begin{center}
          \includegraphics[clip, width=50mm]{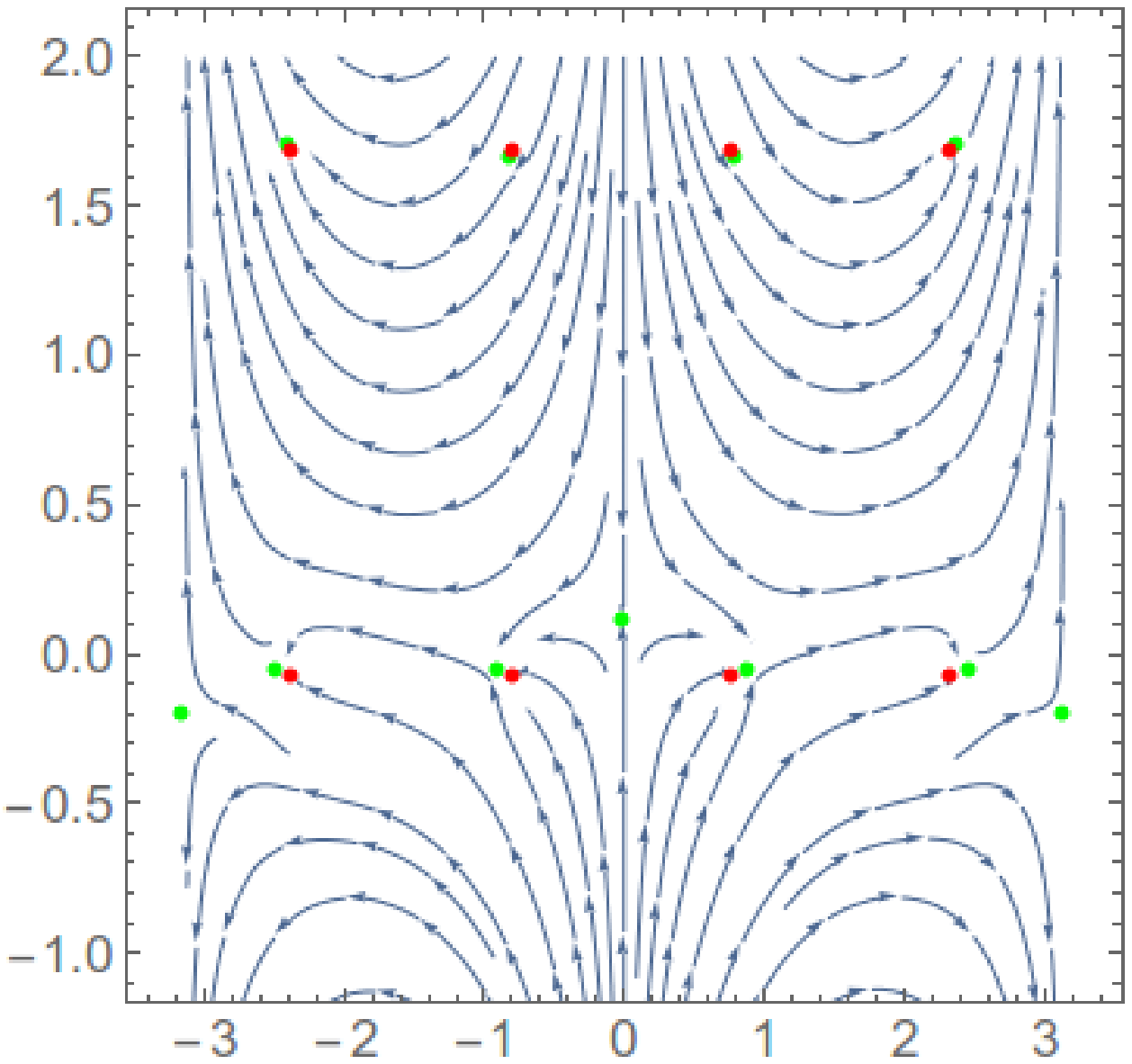}
          \hspace{1.6cm} \\(b) $\mu=0.8$
        \end{center}
      \end{minipage} 
    \end{tabular}
    \caption{Thimble structure in the zero-mode subspace at (a) $\mu=0$ and (b) $\mu=0.8$.
    The green points and red points denote critical points and fermion-zeros, respectively.
    The blue arrows indicate the downward flows.}
    \label{fig:thimble_structure}
  \end{center}
\end{figure}

Here, we apply the Lefschetz thimble method to the Thirring model.
We firstly extend the real variables $A_{n}$ to complex numbers, $A_{n} \rightarrow z_{n}$.
Note that the originally compact variables $e^{i A_{n}}$ become non-compact after this extension.
Then, we solve the complexified saddle point equation,
\be
\frac{\pd S[z]}{\pd z_{n}} =0 \qquad \mbox{for all } n,
\ee
to find out the critical points.
We denote a set of critical points as $\{ \sigma \}$.
The Lefschetz thimble ${\cal J}_\sigma$ associated to a critical point $\sigma$ is defined 
as a union of the downward flow in the complexified configuration space:
\be
\frac{d z_{n}}{dt} = \frac{\pd \bar{S}[\bar{z}]}{\pd z_{n}} \qquad \mbox{with } z \rightarrow \sigma \quad \mbox{as} \quad t \rightarrow - \infty.
\ee
Along a flow line the imaginary part of the action ${\rm Im} S$ remains constant.
The downward flows extend to the region where $h=-{\rm Re}S$ becomes $-\infty$.
Importantly,
since the action diverges at the zero points of the fermion determinant,
\be
\det D[z] =0, \label{eq:zero-mode}
\ee
the thimbles generally can end at those fermion-zeros.
Indeed the zero point condition~(\ref{eq:zero-mode}) can be solved exactly.
When we restrict that $z_{n}$ is independent of $n$, the solutions are written
as
\be
z_{{\rm zero}} = i (\mu \mp  \hat{m}) + \frac{2 \ell+1}{L} \pi, \qquad  \ell \in {\mathbb Z}_{L}.
\ee
Figure~\ref{fig:thimble_structure} shows the thimble structure within the zero-mode subspace at $\mu=0$ and $0.8$.
For $\mu=0$ the downward flows emanating from the origin run into the other critical points at $z=\pm \pi$,
while for $\mu=0.8$ they flow into the determinant zeros.
The thimble structure depends on parameters in the theory.

To decompose the original integration contour into the thimbles,
we need to select the contributing thimbles, which have the
nonzero intersection numbers determined by the dual cycles ${\cal K}_\sigma$ 
sharing the same critical point $\sigma$.
The dual cycle ${\cal K}_\sigma$ is defined by the following equation,
\be
\frac{d z_{n}}{dt} = \frac{\pd \bar{S}[\bar{z}]}{\pd z_{n}} \qquad \mbox{with } z \rightarrow \sigma \quad \mbox{as} \quad t \rightarrow + \infty.
\ee
${\cal K}_\sigma$ emanates from either of the points $\pm i \infty$, where $h=+\infty$.
Roughly, the intersection number $n_\sigma$ is unity (zero) if its dual cycle ${\cal K}_\sigma$ 
does (not) intersect with the original contour.
In general it is hard to determine all intersection numbers because we need whole knowledge on the global thimble structure in the complexified configuration space.
Furthermore, the set of thimbles which contribute to the partition function changes depending on $\mu$.
In the next section, we study the Stokes phenomenon to obtain information of change of intersection numbers.

\section{Stokes phenomenon and intersection numbers}
Here we study the Stokes phenomenon which provides information of change of intersection numbers.
In order for an intersection number to change (with increasing $\mu$), 
the endpoint of ${\cal K}_\sigma$ must change, e.g.,
from $-i\infty$ to $+i\infty$. Inbetween there is a critical value $\mu^*$ 
at which ${\cal K}_\sigma$ is connected to another critical point $\sigma'$.
At this $\mu^*$, ${\cal K}_\sigma$ is overlapping with ${\cal J}_{\sigma'}$.
When two critical points are connected by a flow, we say that the Stokes phenomenon occurs.
A necessary condition for the Stokes phenomenon is as follows:
\be
{\rm Im}S_{\sigma} = {\rm Im}S_{\sigma^{\prime}} +  2 \pi k,
\ee 
with $k \in {\mathbb Z}$.
In ordinary cases such as $\lambda \phi^{4}$ theory the second term on the r.h.s.
is not necessary, but it is needed in our case because of the branch cuts
caused by the $\log \det D$ term in the action.
\begin{figure}[thbp]
  \begin{center}
    \begin{tabular}{cc}
      \begin{minipage}{0.5\hsize}
        \begin{center}
          \includegraphics[clip, width=50mm]{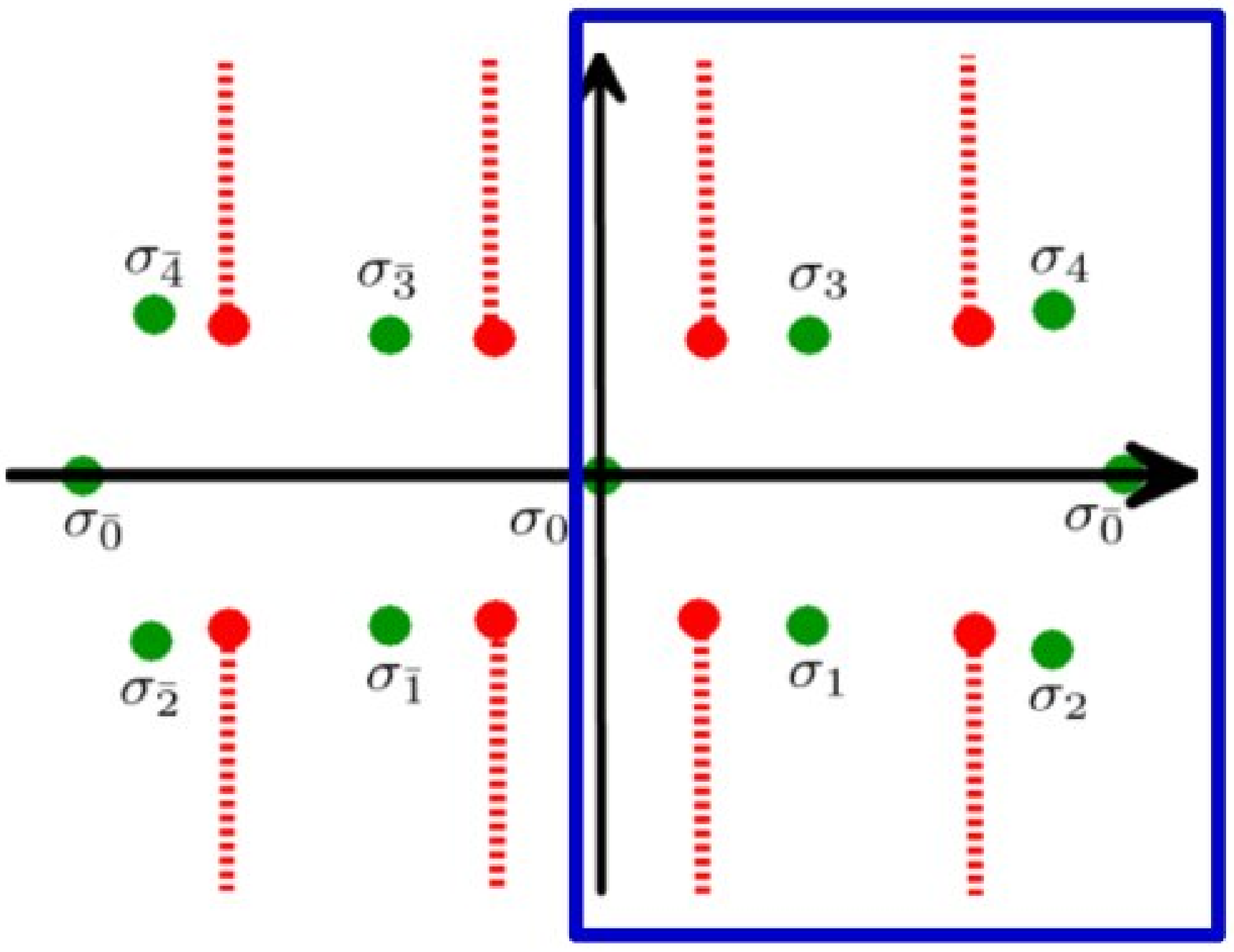}
          \hspace{1.6cm} (a) Critical points and fermion-zeros
        \end{center}
      \end{minipage} 
      \begin{minipage}{0.5\hsize}
        \begin{center}
          \includegraphics[clip, width=55mm]{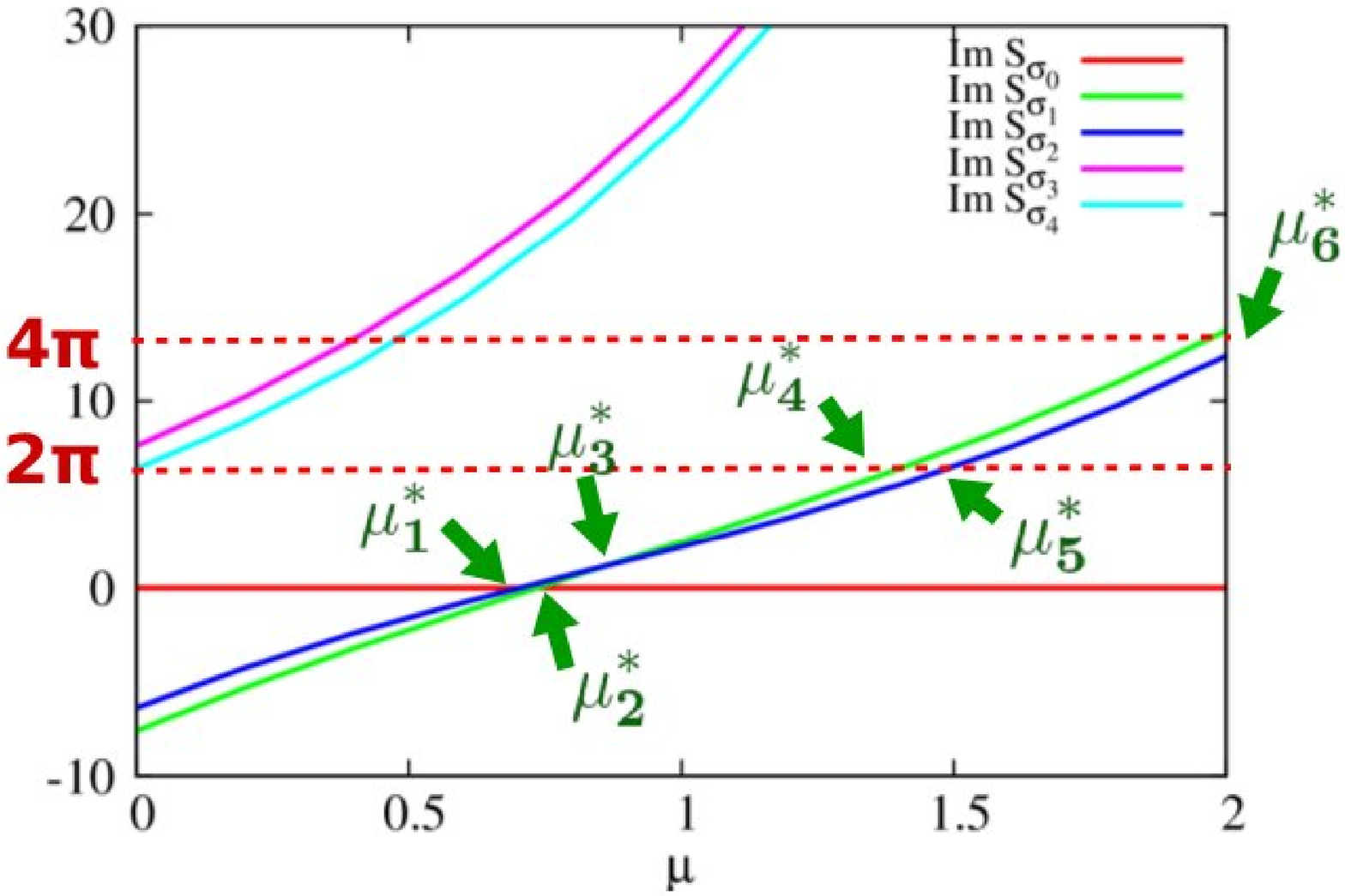}
          \hspace{1.6cm} (b) values of ${\rm Im}S_{\sigma}$ 
        \end{center}
      \end{minipage} 
    \end{tabular}
    \caption{
Critical points and determinant zeros at $L=4,N_{f}=1,\beta=3,m=1$ and values of ${\rm Im}S_{\sigma}$ at those critical points.
    In the left panel, the green and red circles are critical points and fermion-zeros, respectively.
    The red lines running from the fermion-zeros denote branch cuts caused by $\log \det D$ term in the action.
    We investigate the thimble structure of the right-half plane surrounded by a blue square below.
    In the right panel, $\mu^{*}$ denotes the value at which the Stokes phenomenon occurs.
}
    \label{fig:Stokes_mu}
  \end{center}
\end{figure}
Figure~\ref{fig:Stokes_mu} shows the critical points $\sigma_i$ and zero points in the left panel,
and the values of ${\rm Im}S_{\sigma}$ on the critical points in the zero-mode subspace
with $L=4,N_{f}=1,\beta=3,m=1$ in the right panel.
For this parameter set, there are six $\mu^{*}$ at which the Stokes phenomenon occurs.
\begin{figure}[thbp]
  \begin{center}
    \begin{tabular}{ccc}
      \begin{minipage}{0.3333333\hsize}
        \begin{center}
          \includegraphics[clip, width=40mm]{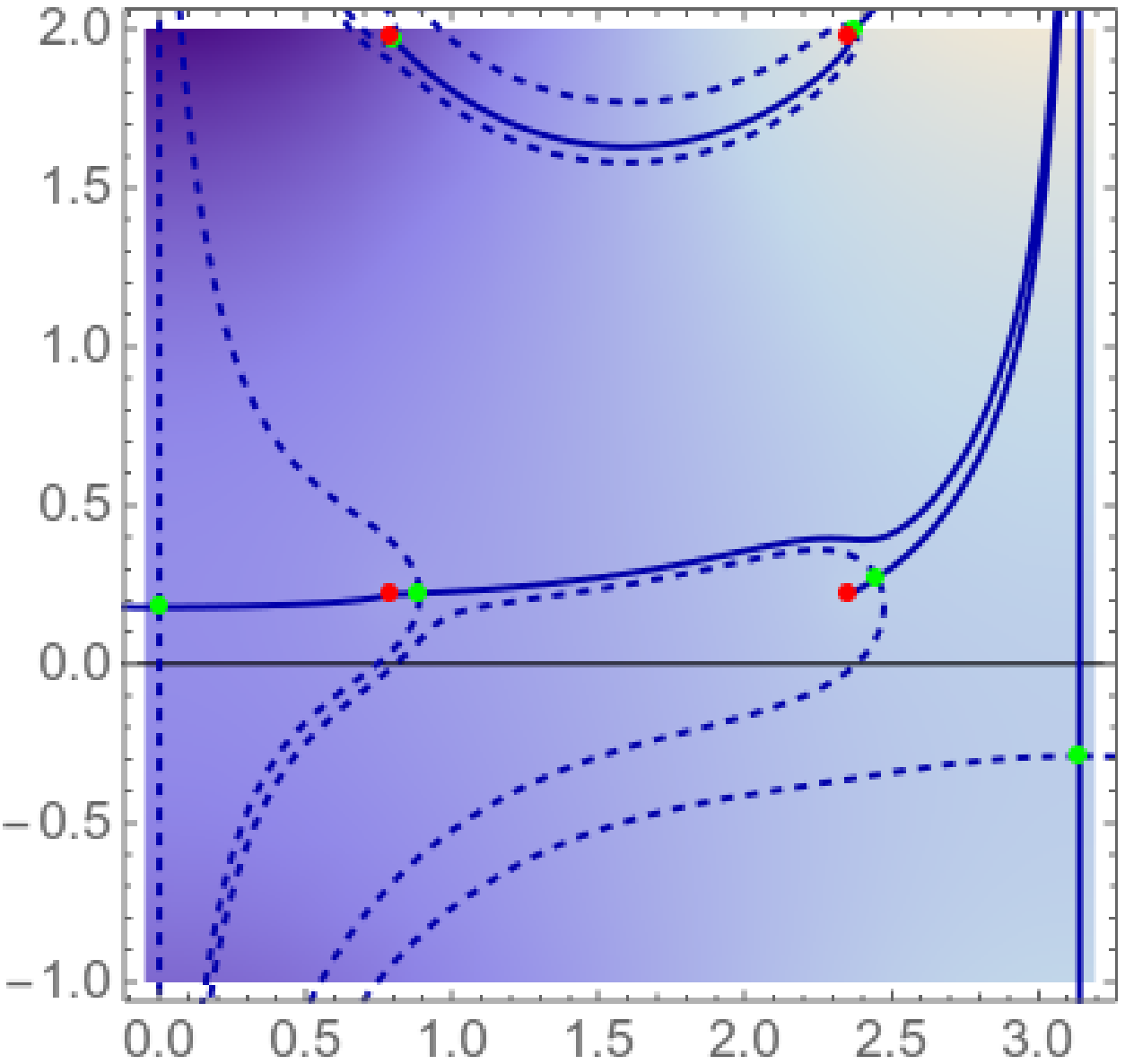}
          \hspace{1.6cm} (a) $\mu<\mu^{*}$
        \end{center}
      \end{minipage} 
      \begin{minipage}{0.3333333\hsize}
        \begin{center}
          \includegraphics[clip, width=40mm]{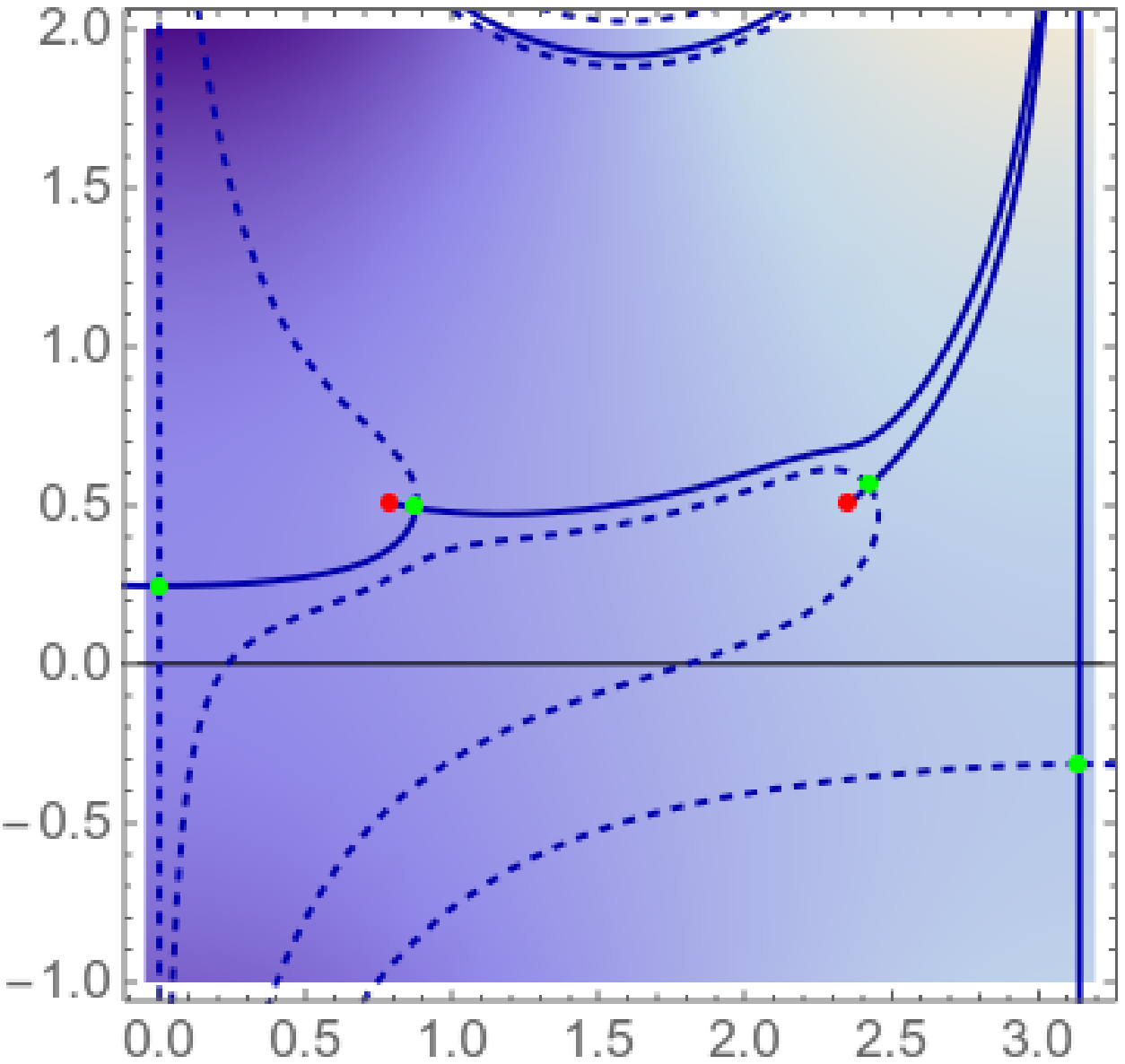}
          \hspace{1.6cm} (b) $\mu=\mu^{*}$
        \end{center}
      \end{minipage} 
      \begin{minipage}{0.3333333\hsize}
        \begin{center}
          \includegraphics[clip, width=40mm]{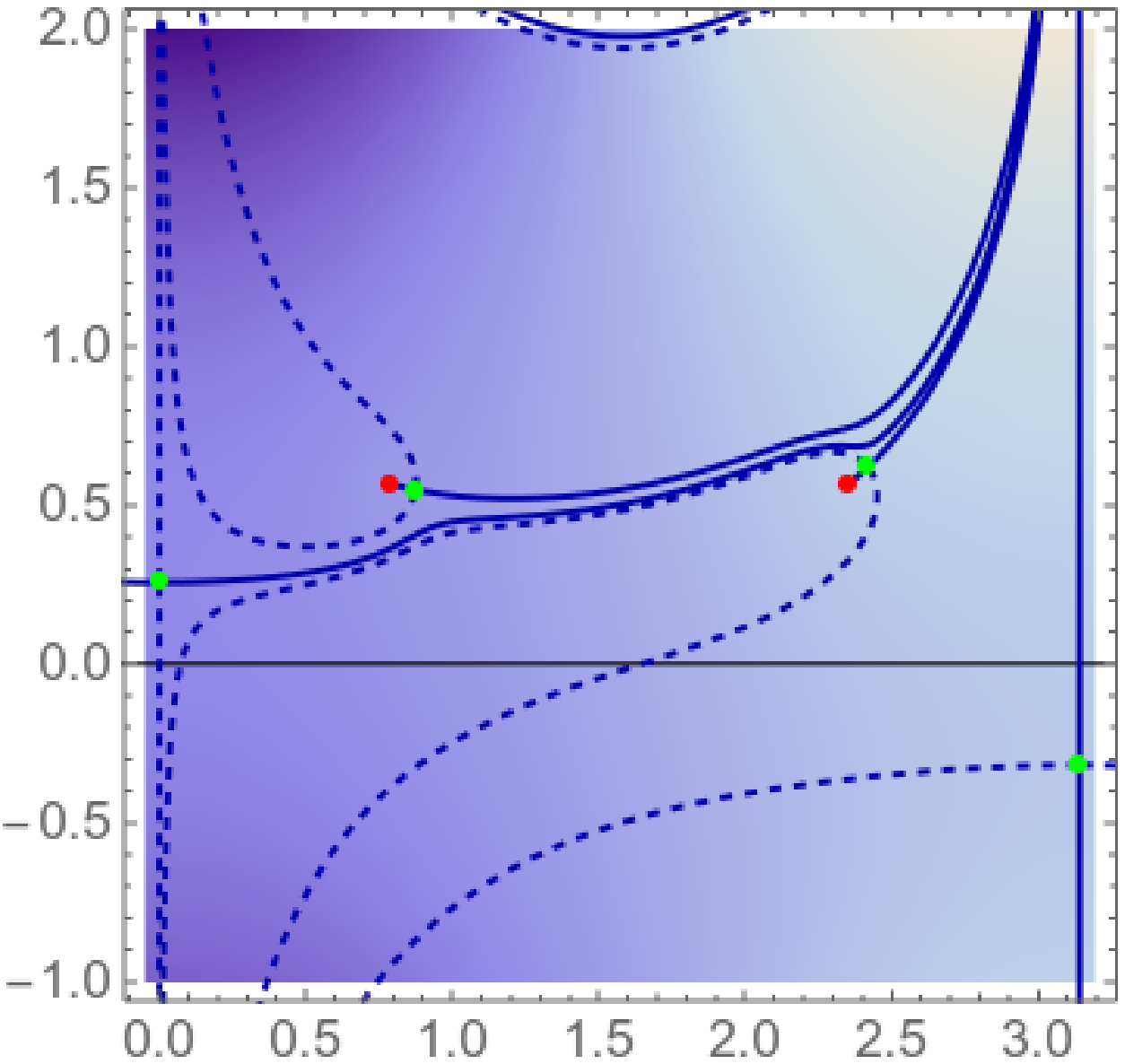}
          \hspace{1.6cm} (c) $\mu>\mu^{*}$
        \end{center}
      \end{minipage} 
    \end{tabular}
    \caption{Lefschetz thimble structure around some $\mu_4^{*}$.
    The green and red points denote critical points and fermion-zeros, respectively.
    The solid (dotted) lines show the thimbles (dual cycles).}
    \label{fig:Stokes}
  \end{center}
\end{figure}
Figure~\ref{fig:Stokes} shows the thimble structures around $\mu_4^{*}$.
The dual cycle ${\cal K}_{\sigma_1}$ intersects
the real axis at $\mu<\mu^{*}$ but it does not at $\mu>\mu^{*}$. 
Hence, $\mu_4^{*}$ is the critical value for change of the intersection number $n_{\sigma_1}$.
\begin{figure}[thbp]
  \begin{center}
    \begin{tabular}{ccc}
      \begin{minipage}{0.3333333\hsize}
        \begin{center}
          \includegraphics[clip, width=40mm]{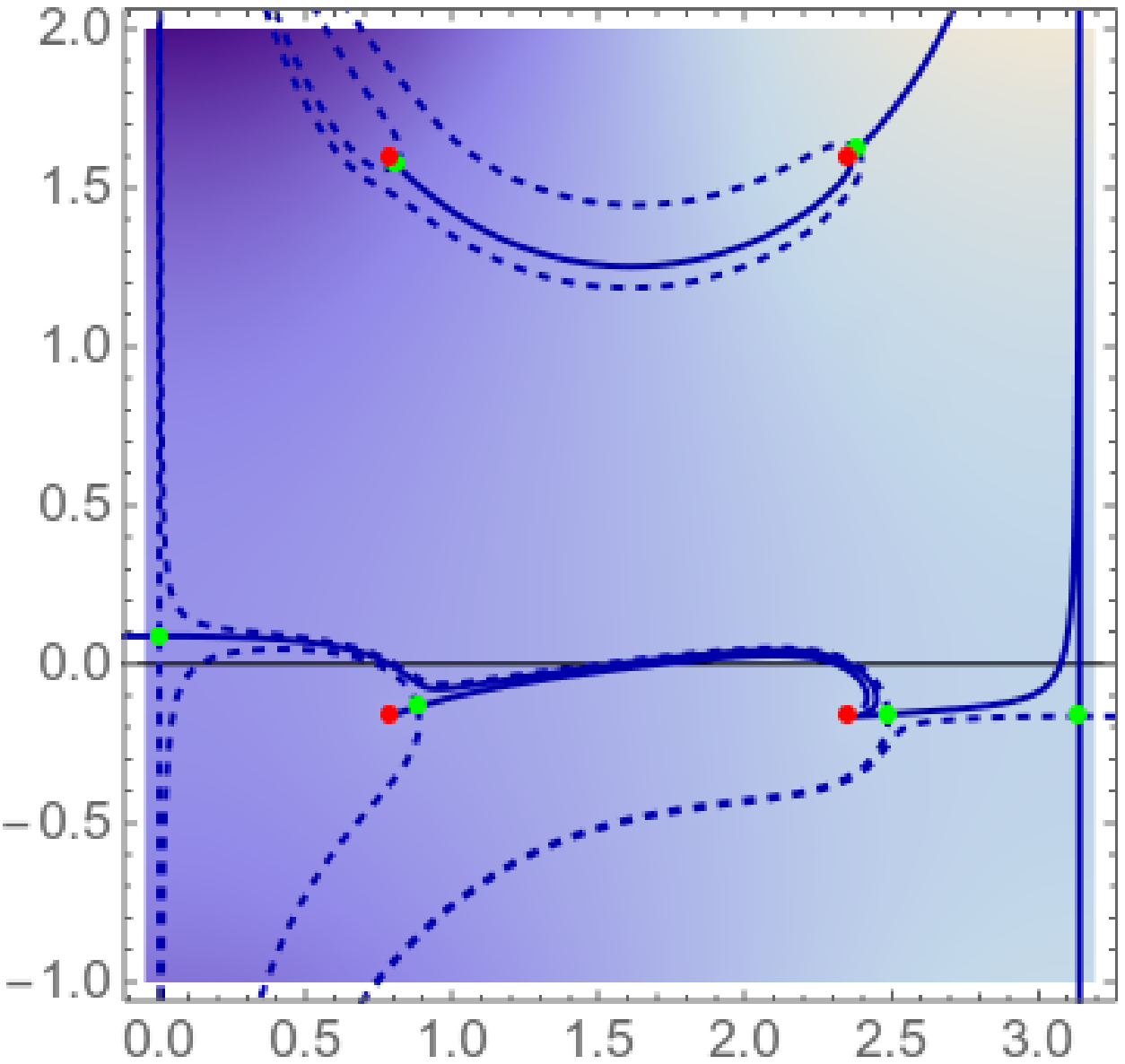}
          \hspace{1.6cm} (a) $\mu^{*}_{1}<\mu<\mu^{*}_{2}$
        \end{center}
      \end{minipage} 
      \begin{minipage}{0.3333333\hsize}
        \begin{center}
          \includegraphics[clip, width=40mm]{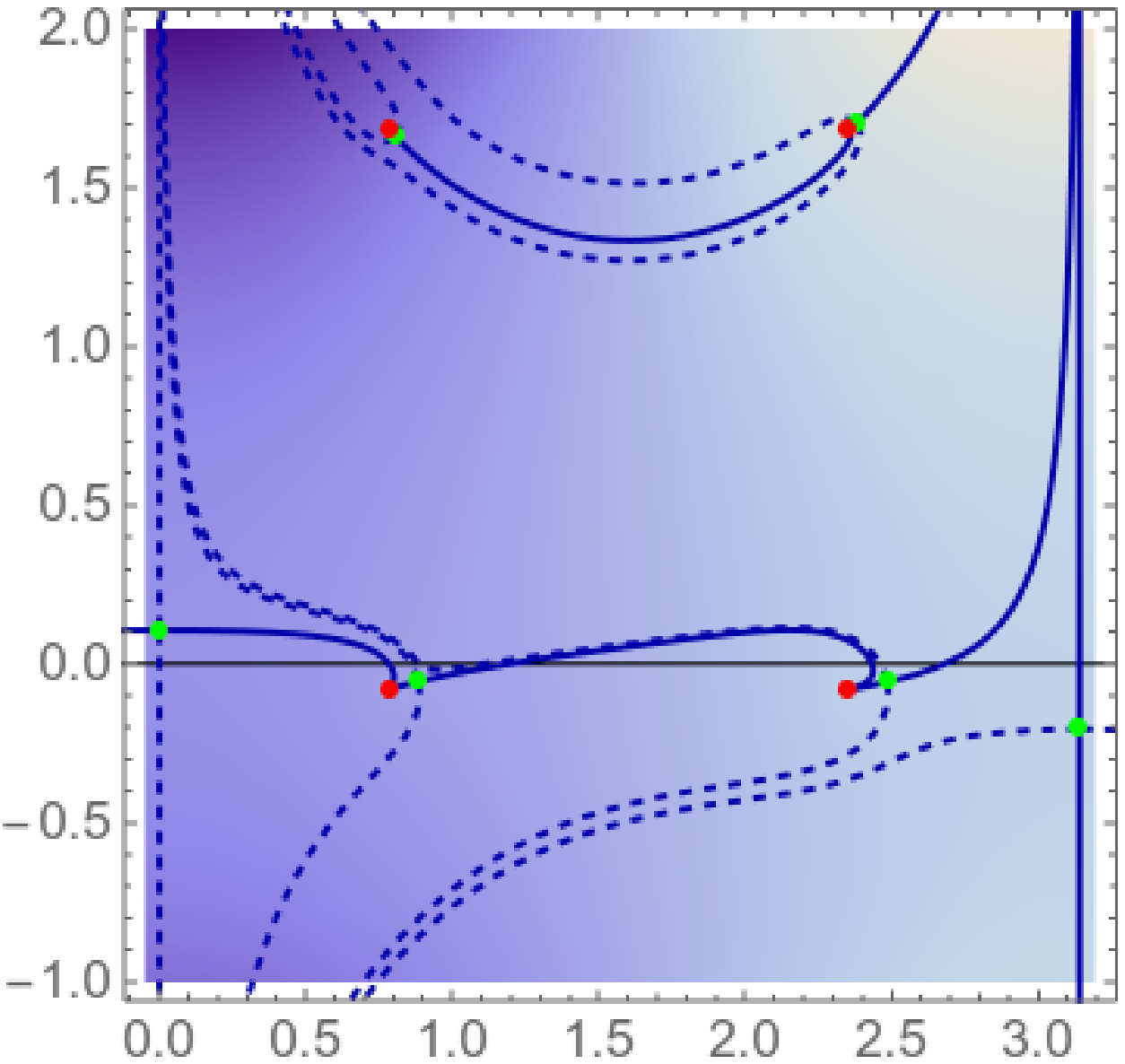}
          \hspace{1.6cm} (b) $\mu^{*}_{2}<\mu<\mu^{*}_{3}$
        \end{center}
      \end{minipage} 
      \begin{minipage}{0.3333333\hsize}
        \begin{center}
          \includegraphics[clip, width=40mm]{./figs/MT_thimble34.eps}
          \hspace{1.6cm} (c) $\mu^{*}_{3}<\mu<\mu^{*}_{4}$
        \end{center}
      \end{minipage} \\
      \begin{minipage}{0.3333333\hsize}
        \begin{center}
          \includegraphics[clip, width=40mm]{./figs/MT_thimble45.eps}
          \hspace{1.6cm} (d) $\mu^{*}_{4}<\mu<\mu^{*}_{5}$
        \end{center}
      \end{minipage} 
      \begin{minipage}{0.3333333\hsize}
        \begin{center}
          \includegraphics[clip, width=40mm]{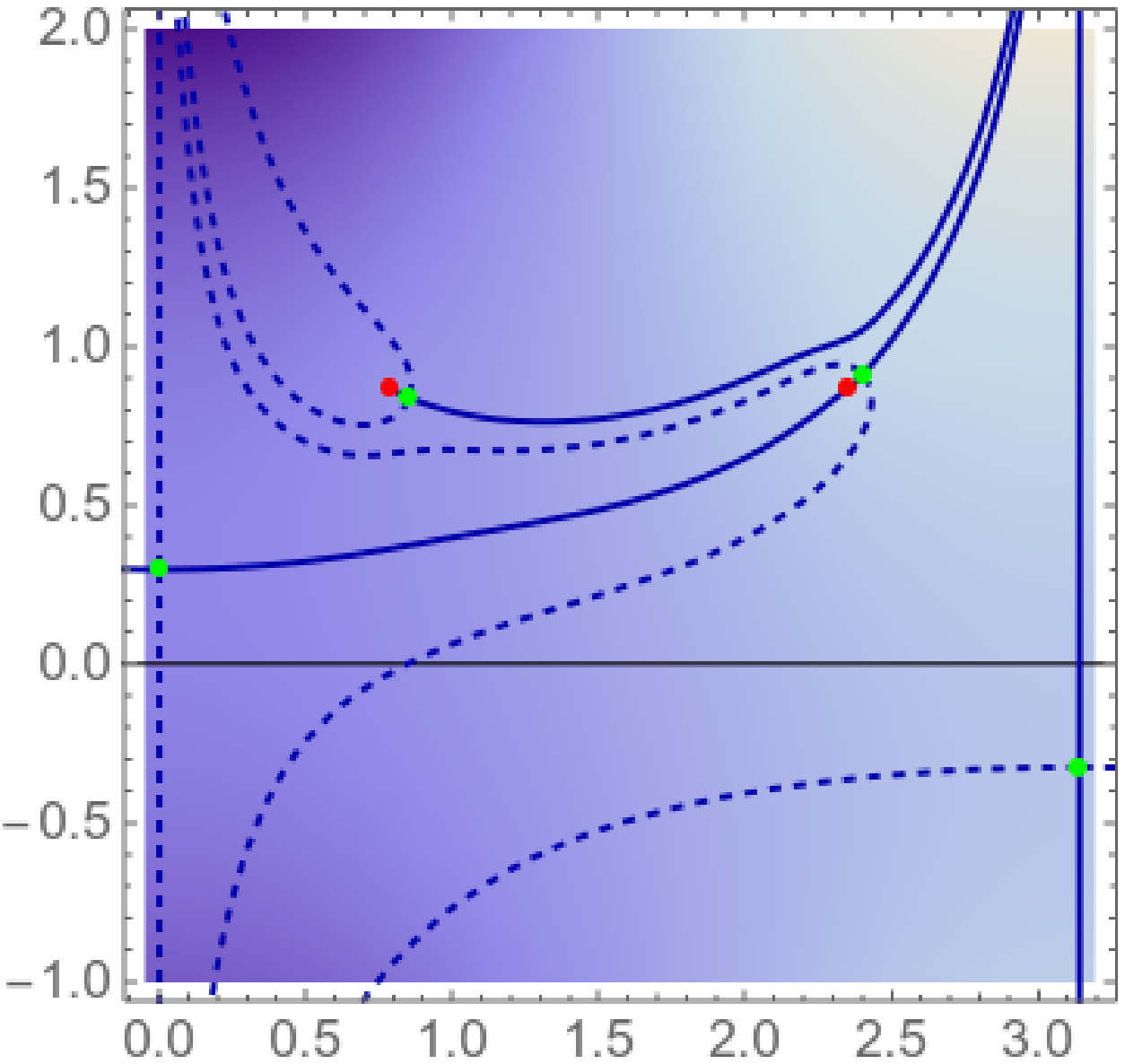}
          \hspace{1.6cm} (e) $\mu^{*}_{5}<\mu<\mu^{*}_{6}$
        \end{center}
      \end{minipage} 
      \begin{minipage}{0.3333333\hsize}
        \begin{center}
          \includegraphics[clip, width=40mm]{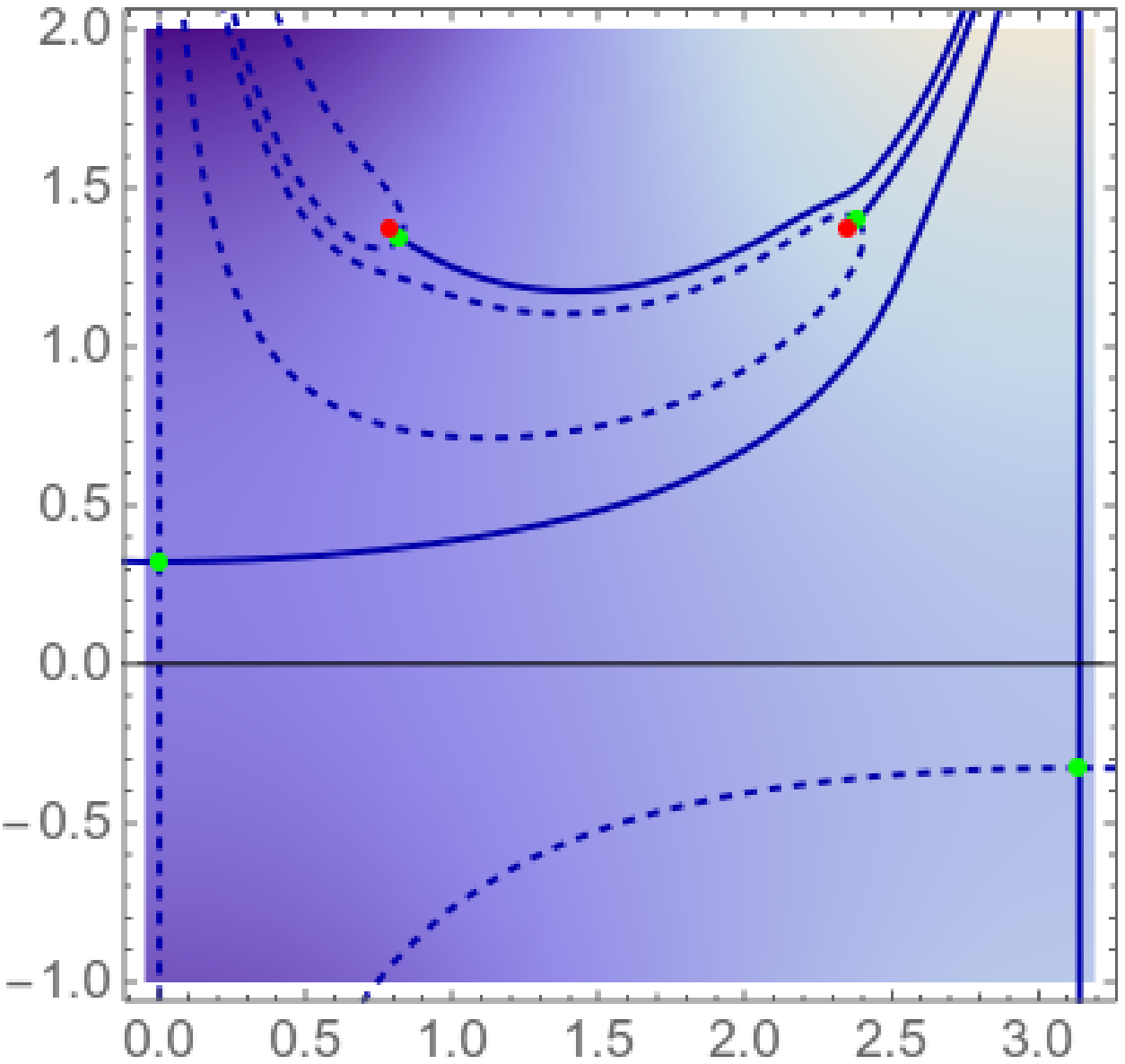}
          \hspace{1.6cm} (f) $\mu^{*}_{6}<\mu$
        \end{center}
      \end{minipage} 
    \end{tabular}
    \caption{
$\mu$-dependence of the thimble structure in the right-half plain of the zero-mode configuration space at $L=4,N_{f}=1,\beta=3,m=1$.
The contour of the zero-mode partition function can be constructed as follows:
(a)${\cal C} = {\cal J}_{\sigma_{0}}+{\cal J}_{\sigma_{2}}+{\cal J}_{\sigma_{\bar{2}}}$, (b)${\cal C}= {\cal J}_{\sigma_{0}}+{\cal J}_{\sigma_{2}}+{\cal J}_{\sigma_{\bar{2}}}$, (c)${\cal C}= {\cal J}_{\sigma_{0}}+{\cal J}_{\sigma_{1}}+{\cal J}_{\sigma_{\bar{1}}}$, (d)${\cal C}= {\cal J}_{\sigma_{0}}$, (e)${\cal C}= {\cal J}_{\sigma_{0}}+{\cal J}_{\sigma_{2}}+{\cal J}_{\sigma_{\bar{2}}}$, (f)${\cal C}= {\cal J}_{\sigma_{0}}$.
}
    \label{fig:mu-dep_Stokes}
  \end{center}
\end{figure}
We can construct the integration contour of the zero-mode partition function by thimbles.
In Fig.~\ref{fig:mu-dep_Stokes}, we present the right-half plain of the zero-mode configuration space at several values of $\mu$.
From this figure, we find that the thimble ${\cal J}_{\sigma_{0}}$ always contributes to the partition function.
Actually, by comparing ${\rm Re}S$ at $\sigma_i$, this thimble ${\cal J}_{\sigma_0}$ is the most dominant.
One of the important questions then is how accurately we can reproduce the exact solutions only by 
this dominant thimble ${\cal J}_{\sigma_0}$.

\section{HMC simulation}
\begin{figure}[thbp]
  \begin{center}
    \begin{tabular}{cc}
      \begin{minipage}{0.5\hsize}
        \begin{center}
          \includegraphics[clip, width=60mm]{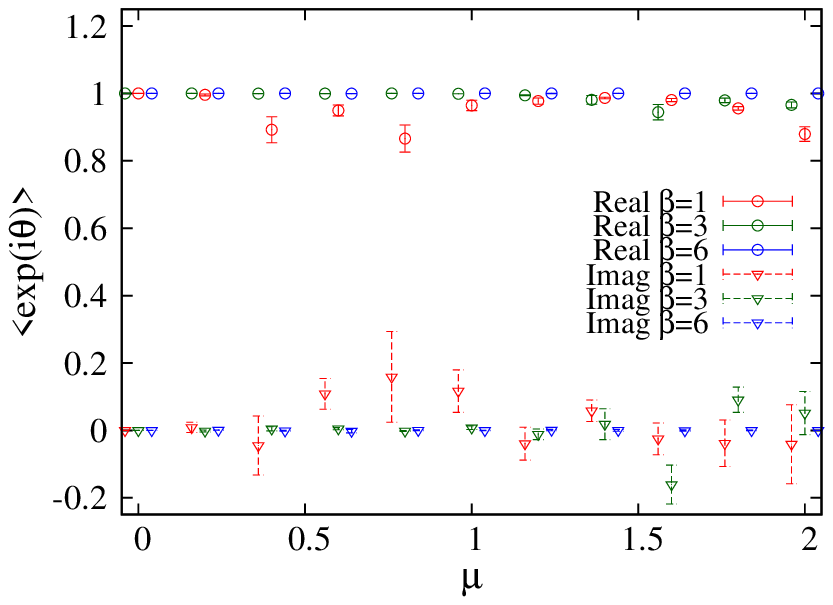}
          \hspace{1.6cm} (a) $L=4$
        \end{center}
      \end{minipage} 
      \begin{minipage}{0.5\hsize}
        \begin{center}
          \includegraphics[clip, width=60mm]{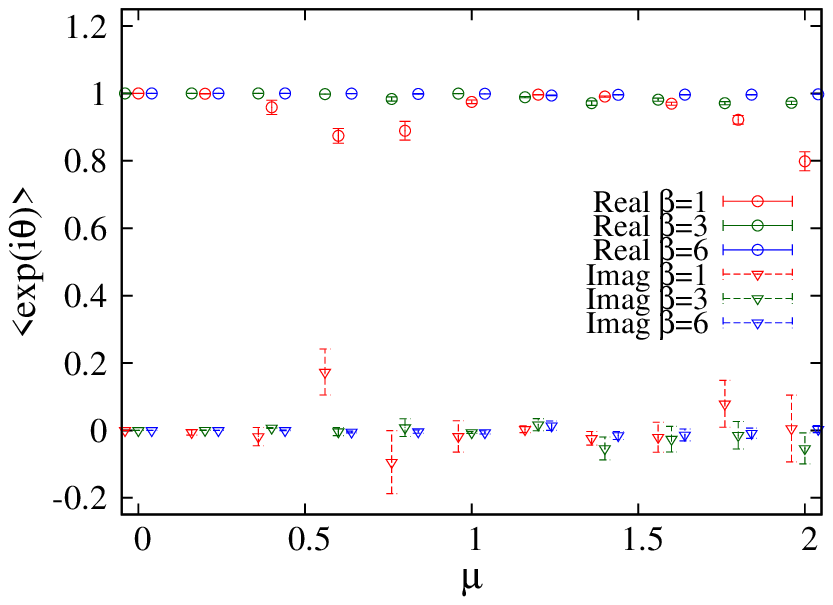}
          \hspace{1.6cm} (b) $L=8$
        \end{center}
      \end{minipage} 
    \end{tabular}
    \caption{Residual phase at (a) $L=4$ and (b) $L=8$.
      The fermion mass is fixed at $m=1$.
      The red, green and blue colors denote the results at $\beta=1,3$ and $6$, respectively.
      The real (imaginary) part is denoted by $\circ$ ($\triangledown$).
}
    \label{fig:phase}
  \end{center}
\end{figure}

\begin{figure}[htbp]
  \begin{center}
    \begin{tabular}{cc}
      \begin{minipage}{0.5\hsize}
        \begin{center}
          \includegraphics[clip, width=60mm]{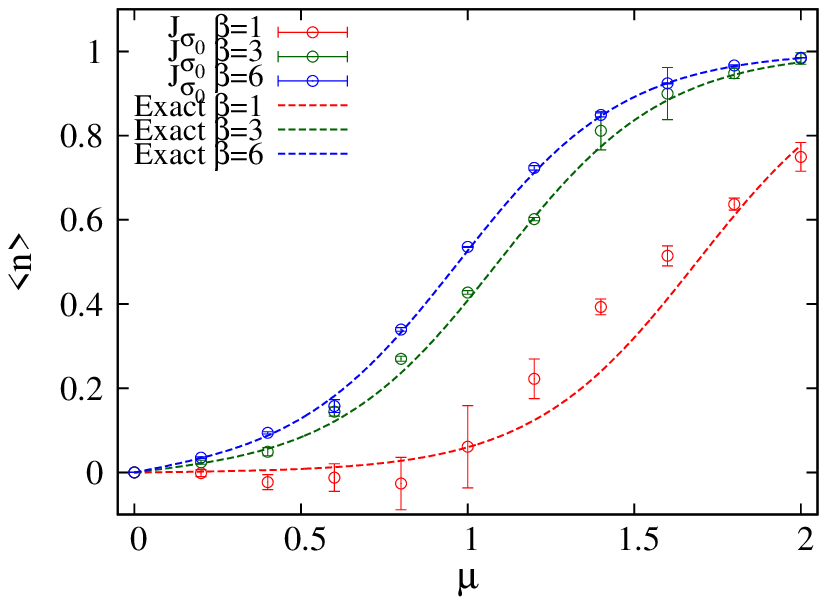}
          \hspace{1.6cm} (a) $L=4$
        \end{center}
      \end{minipage} 
      \begin{minipage}{0.5\hsize}
        \begin{center}
          \includegraphics[clip, width=60mm]{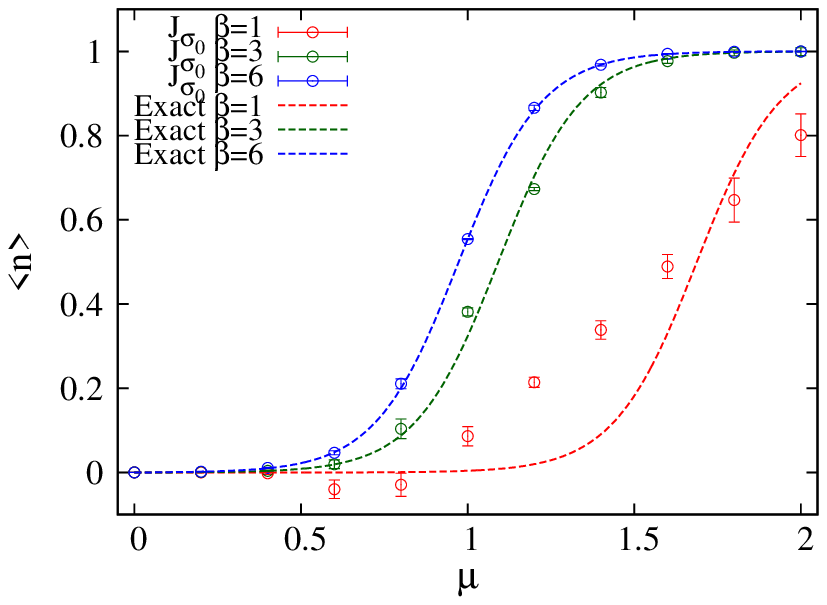}
          \hspace{1.6cm} (b) $L=8$
        \end{center}
      \end{minipage} 
    \end{tabular}
    \caption{$\mu$-dependence of the number density at (a)$L=4$ and (b)$L=8$ with a unit fermion mass.
      The squared coupling constant is chosen as $\beta=1$ (red), $3$ (green), and $6$ (blue).
      We represent numerical and exact results using colored points and solid lines, respectively.
    }
    \label{fig:number}
  \end{center}
\end{figure}

\begin{figure}[thbp]
  \begin{center}
    \begin{tabular}{cc}
      \begin{minipage}{0.5\hsize}
        \begin{center}
          \includegraphics[clip, width=60mm]{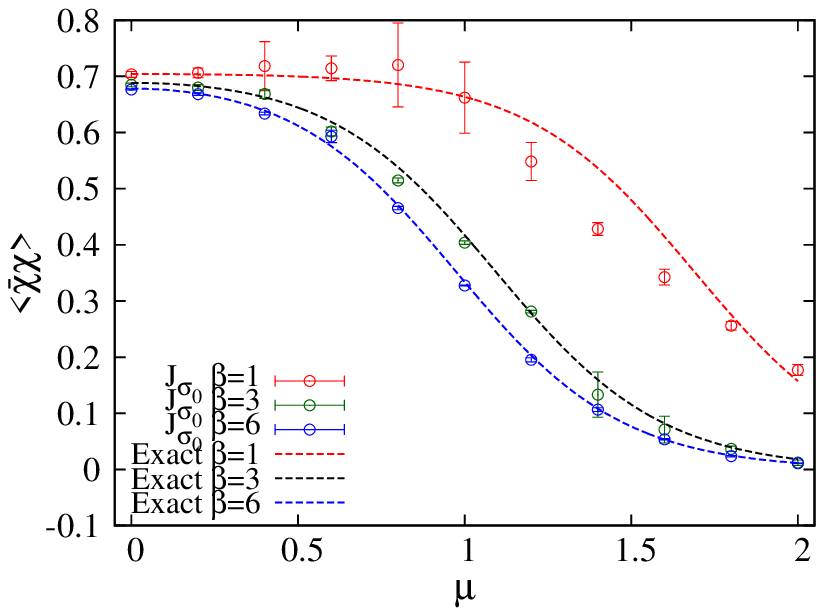}
          \hspace{1.6cm} (a) $L=4$
        \end{center}
      \end{minipage} 
      \begin{minipage}{0.5\hsize}
        \begin{center}
          \includegraphics[clip, width=60mm]{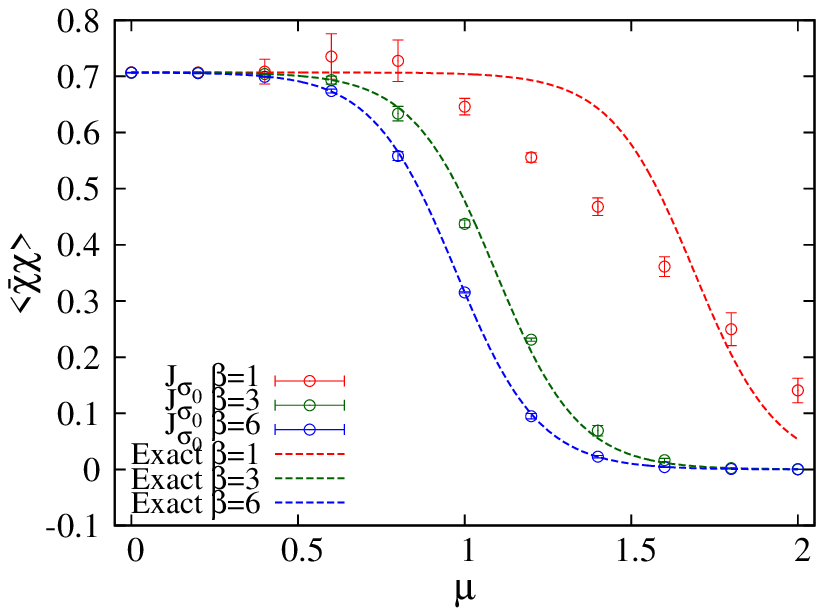}
          \hspace{1.6cm} (b) $L=8$
        \end{center}
      \end{minipage} 
    \end{tabular}
    \caption{$\mu$-dependence of the fermion condensation at (a)$L=4$ and (b)$L=8$ with a unit fermion mass.
      The squared coupling constant is chosen as $\beta=1$ (red), $3$ (green), and $6$ (blue).
      We represent numerical and exact results using colored points and solid lines, respectively.
    }
    \label{fig:meson}
  \end{center}
\end{figure}

We perform lattice simulations on the single thimble ${\cal J}_{0}$ 
in order to test how accurately the dominant thimble can reproduce the exact solutions.
We have employed  the HMC algorithm proposed in Ref.\cite{Fujii:2013sra} with a few
improvements by 
introducing a scale parameter $\lambda$ to rescale the variables as $z_{n} \rightarrow \lambda z_{n}$ 
and by using the adaptive step size in the Runge-Kutta method.
Figure~\ref{fig:phase} shows the residual phase averages 
at $L=4,8$,  $\beta=1,3,6$ and $m=1$. 
At those parameters, the averages of the residual phase 
stabilize around $\langle \exp(i \theta) \rangle \sim 0.8$-$1$.
Hence, we can rely on the phase reweighting to estimate observables.
Numerical results for the number density and the fermion condensation 
are plotted in Figs.~\ref{fig:number} and \ref{fig:meson}.
At $\beta=3,6$, the numerical results are in good agreement with the exact solutions within the $\mu$-range considered.
However, at $\beta=1$, we find discrepancies between the numerical results and the exact ones in the crossover region.

%


\section{Summary}
We have applied the Lefschetz thimble method to the (0+1) dimensional Thirring at finite density.
We firstly investigate thimble structure, and then study the Stokes phenomenon to identify the contributing thimbles to the evaluation of the partition function within the $n$-independent complex field subspace.
We next performed HMC simulations on the dominant thimble ${\cal J}_{\sigma_0}$, which ends at the determinant-zeros
at finite chemical potential $\mu$.
At  large $\beta$ the simulation results reproduce the exact one very well, 
but at small $\beta$ we have observed the discrepancies in the crossover region of $\mu$.
This numerical result is consistent with the analytical study of  the model's thimble structure
and shows that 
in the crossover region
we need to take into account the contributions from the subdominant thimbles 
to reproduce the exact results appropriately.


\begin{thebibliography}{99}

\bibitem{deForcrand:2010ys} 
  P.~de Forcrand,
  PoS LAT {\bf 2009}, 010 (2009)
  [arXiv:1005.0539 [hep-lat]].

\bibitem{Sexty:2014dxa} 
  D.~Sexty,
  PoS LATTICE {\bf 2014}, 016 (2014)
  [arXiv:1410.8813 [hep-lat]].


\bibitem{Witten:2010cx} 
  E.~Witten,
  arXiv:1001.2933 [hep-th].

\bibitem{Cristoforetti:2012su}
  M.~Cristoforetti {\it et al.}  [AuroraScience Collaboration],
  Phys.\ Rev.\ D {\bf 86}, 074506 (2012)
  [arXiv:1205.3996 [hep-lat]].

\bibitem{Fujii:2013sra} 
  H.~Fujii, D.~Honda, M.~Kato, Y.~Kikukawa, S.~Komatsu and T.~Sano,
  JHEP {\bf 1310}, 147 (2013)
  [arXiv:1309.4371 [hep-lat]].

\bibitem{Pawlowski:2014ada} 
  J.~M.~Pawlowski, I.~O.~Stamatescu and C.~Zielinski,
  arXiv:1402.6042 [hep-lat].

\bibitem{Fujii:2015bua} 
  H.~Fujii, S.~Kamata and Y.~Kikukawa,
  arXiv:1509.08176 [hep-lat].

\bibitem{Fujii:2015vha} 
  H.~Fujii, S.~Kamata and Y.~Kikukawa,
  arXiv:1509.09141 [hep-lat].
\end{thebibliography}
\end{document}